\documentclass[10pt]{ismd08}
\usepackage{graphicx}
\usepackage{epsfig}

\def\be{\begin{equation}}
\def\ee{\end{equation}}
\begin{document}
\title{Saturation effects in final states due to CCFM with absorptive boundary}
\author{Krzysztof Kutak\protect\footnote{\ \ speaker},
 Hannes Jung}
\institute{$Deutsches\,\,Elektronen-Synchrotron$\\
Notkestr 85, 22603, Hamburg, Germany  
}
\maketitle
\begin{abstract}
We apply the absorptive boundary prescription to include saturation effects in CCFM evolution equation.
We are in particular interested in saturation effects in exclusive processes which can be studied using Monte Carlo event generator CASCADE.
We calculate cross section for three-jet production and distribution of charged hadrons.
\end{abstract}
\vspace{-0.5cm}
\section{Introduction}
At the dawn of LHC it is desirable to have tools which could be safely used to evolve colliding protons to any point of available in collision phase space. 
It is also desirable to have formulation within Monte Carlo framework because this allows to study complete  events (see also contribution of E. Avsar to
ISMD 08).
At present there are two main approaches within pQCD which can be applied to describe evolution of the
parton densities: collinear factorisation with integrated parton densities, with DGLAP as the master equation and $k_T$ factorisation with
unintegrated gluon density with BFKL as the master equation 
\cite{BFKL}.
These two approaches  resum different perturbative series and are  valid in different 
regimes of the longitudinal momentum fraction carried by the partons. However, they tend to merge at higher orders meaning that one is a source of subleading corrections for
the other. The economic way to combine information from both of them is to use the CCFM 
\cite{CCFM} 
approach  which interpolates
between DGLAP and BFKL and  which has the advantage of being applicable to Monte Carlo simulation of final states.
However, if one wants to study physics at largest energies available at LHC one has to go beyond DGLAP, CCFM or BFKL because
all these equations were derived in an
approximation of dilute partonic system where partons do not overlap or to put it differently do not recombine.
Because of this those equations cannot be safely extrapolated towards high
energies, as this is in conflict with unitarity requirements.  
To account for dense partonic systems one has to introduce a mechanism which allows partons to recombine. 
There are various ways to approach this problem 
\cite{unit}, 
here we are interested in the one which can be directly formulated within $k_T$ 
factorisation approach \cite{hautmann}. 
In this approach one can formulate momentum space version \cite{kutak} of the  Balitsky-Kovchegov equation \cite{BK}  
which sums up 
large part of important terms for saturation and which is a nonlinear extension of the BFKL equation.
As it is a nonlinear equation  it is quite cumbersome but one can avoid complications coming from nonlinearity by applying absorptive 
boundary conditions \cite{MT} which mimics the nonlinear term in the BK equation.
Here, in order to have description of exclusive processes and account for saturation effects we use CCFM evolution equation together with 
absorptive boundary implemented in CASCADE Monte Carlo event generator \cite{hannes}.\\
In section one we show description of $F_2$ data using CCFM equation. In section two we describe way to incorporate saturation 
effects.
In section three we show results for angular distribution of three jets and distribution of charged particles. 
\vspace{-0.5cm}
\section{CCFM evolution equation and $F_2$}
\begin{figure}[b!]
\centerline{\epsfig{file=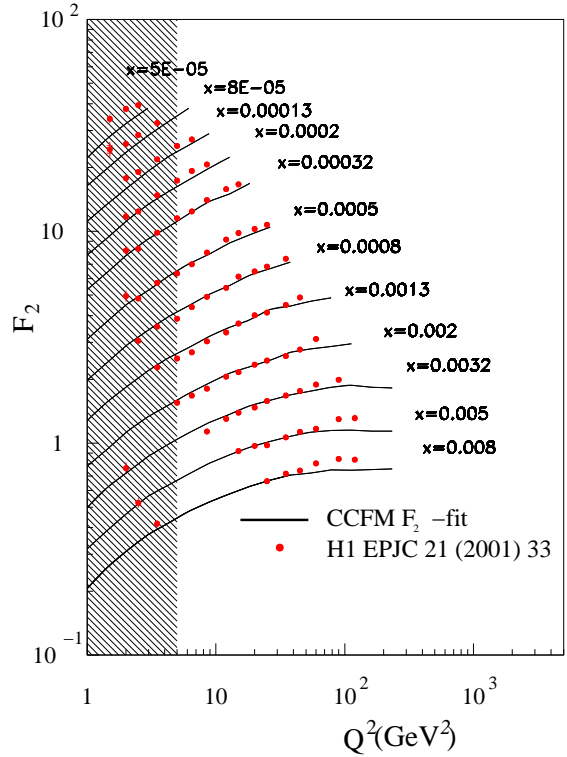,height=10cm}}
\caption{\em $F_2$ description of HERA data with CCFM evolution equation}
\label{fig:F2nosat}
\end{figure}
The CCFM evolution equation is a linear evolution equation which sums up a cascade of gluons under the assumption that gluons are strongly
ordered in an angle of emission.
This can be schematically written as:
$xA(x,k_T^2,q^2)=xA_0(x,k_T^2,q^2)+K\otimes xA(x,k_T^2,q^2)$
where $x$ is the longitudinal momentum fraction of the proton carried by the gluon, $k_T$ is its transverse momentum and $q$ is a factorisation scale.
The initial gluon's distribution  $xA_0(x,k_T,\mu^2)=Nx^{Bg}(1-x)^4exp\left[(k-\mu)^2/\sigma^2\right]$ parameters are to be determined by fit to data.
At present we keep parameters $\mu$ and $\sigma$ fixed and fit $N$ and $Bg$.
Using $k_T$ factorisation theorem gluon density coming from the CCFM equation can be applied to calculate  $F_2$ and compare with measurements.
In the $k_T$ factorisation approach the observables are calculated via convolution of an off-shell hard matrix element with gluon density. The appropriate
formula in schematic form for $F_2$ reads:
$F_2(x,Q^2)=\Phi(x,k^2_T,Q^2)\otimes xA(x,k^2_T,q^2(Q^2))$
where the convolution symbol stands for integration in longitudinal and transversal momenta.
From Fig. \ref{fig:F2nosat} we see agreement with $F_2$ measurements. We should however note
that at the LHC for processes in the forward region  we will probe the gluon density at smaller $x$ than at HERA and unitarity corrections could 
be visible.
\vspace{-0.5cm}
\section{$F_2$ from CCFM with saturation}
\begin{figure}[t!]
  \begin{picture}(490,170)
    \put(-15, -75){
      \includegraphics{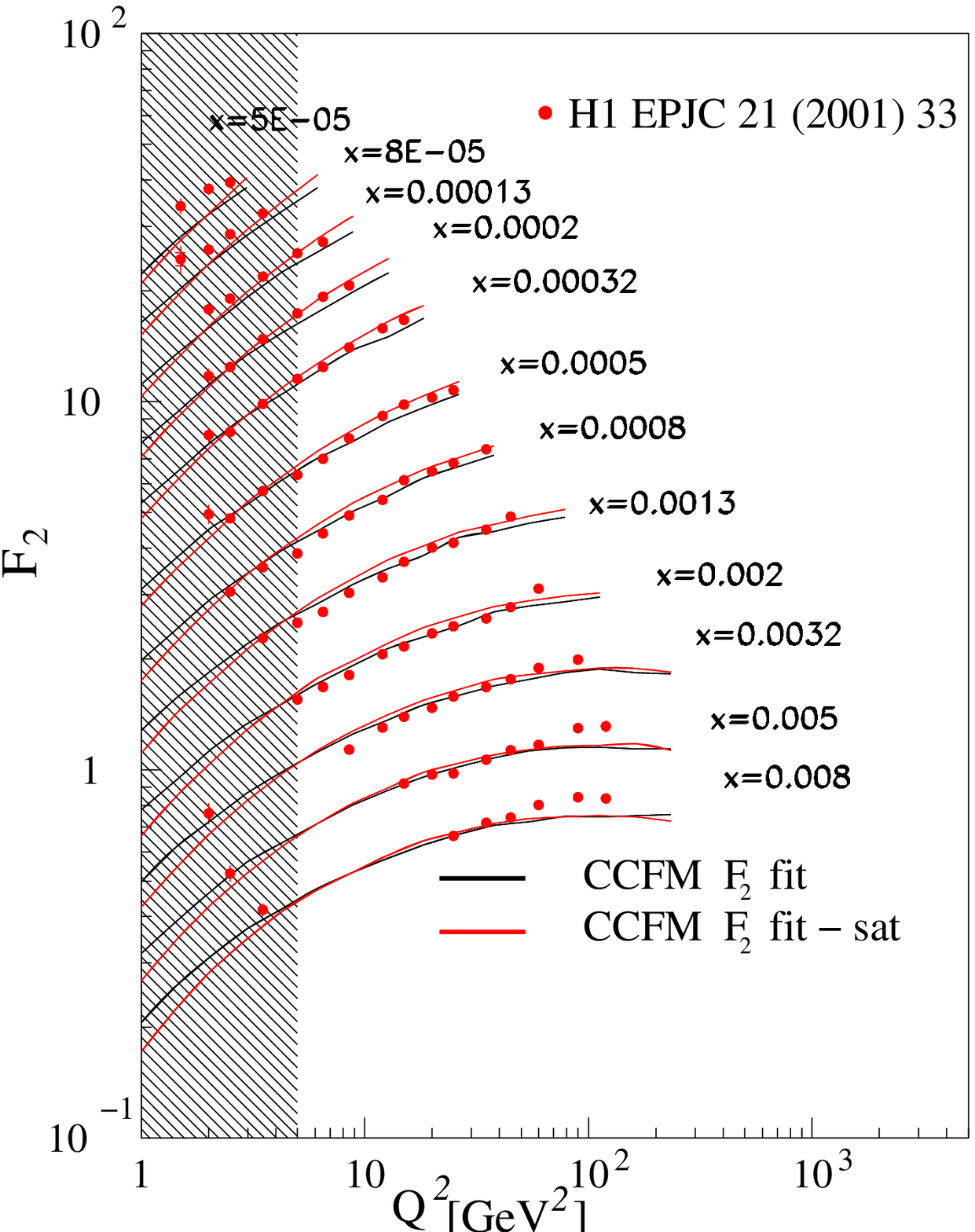}
    }
 \put(180, -45){
      \includegraphics{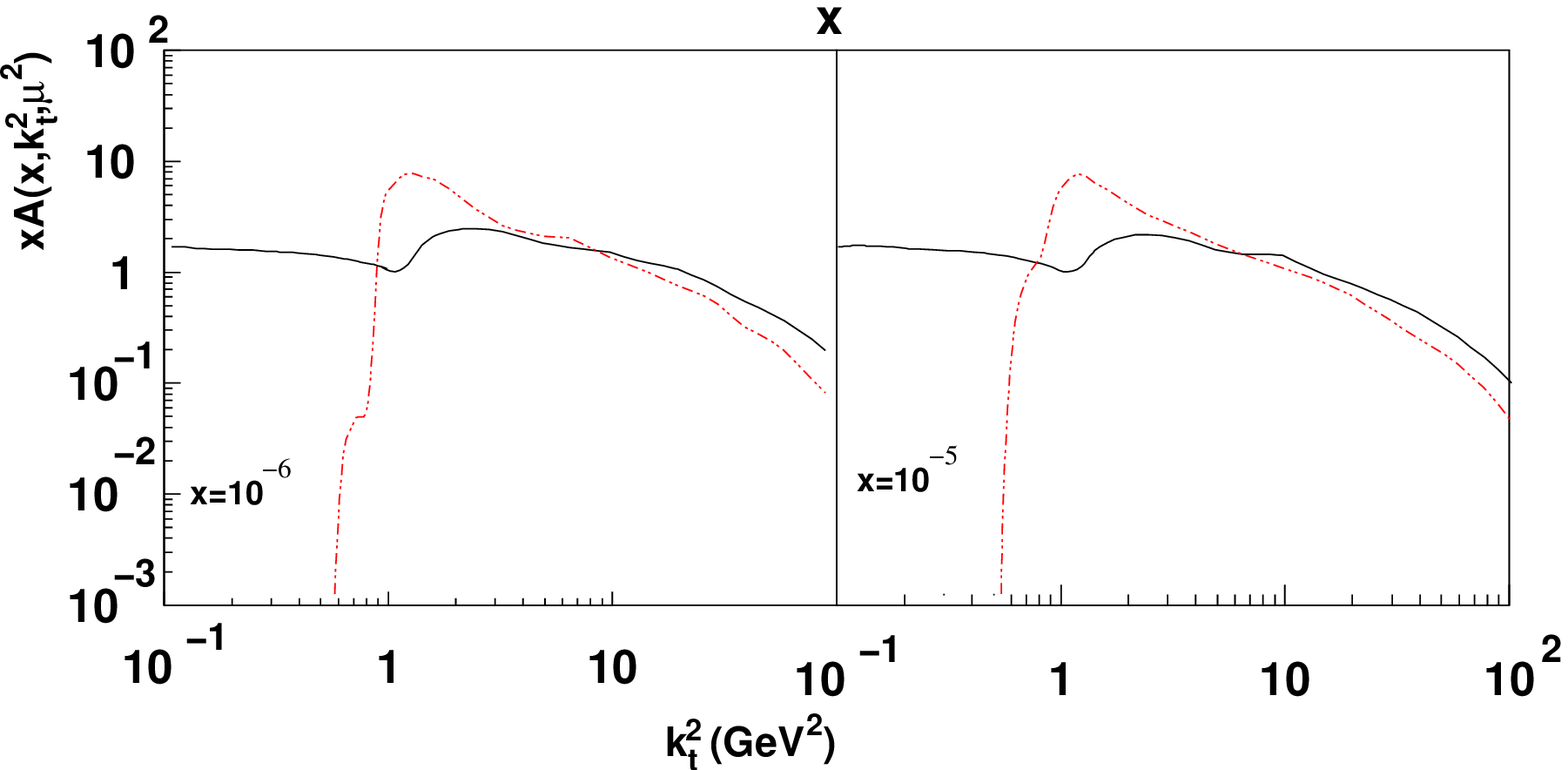}
    }
 \end{picture}
\vspace{1.7cm}
\caption{\em (left) $F_2$ calculated using CCFM with saturation compared to CCFM and to the data. (right) Comparison of gluon density obtained 
from CCFM with saturation to gluon density from CCFM as a function of $k_T^2$ for $x=10^{-5}$, $x=10^{-6}$ }
\label{fig:F2sat}
\end{figure}
The CCFM equation predicts the gluon density which behaves like $A(x,k^2,\mu^2)\sim x^{\beta}$
and this power like behaviour is in conflict with unitarity bounds. 
As it has been already stated the way to introduce part of unitarity corrections is to introduce nonlinear terms to the BFKL or CCFM evolution equation.
The nonlinearity gives rise to the so called energy dependent saturation scale below which gluon density is suppressed. 
Following an idea of A. Mueller and D. Triantafyllopoulos  we model the saturation effects by introducing an absorptive
 boundary which mimics the nonlinear term. In the original approach it was
 required that the BFKL amplitude  should be equal to
unity for a certain combination of $k_T^2$ and $x$. Here we introduce the energy dependent cutoff on transverse gluon momenta which acts as 
absorptive boundary and slows down the rate of growth of the gluon density. As a prescription for the cutoff we use the GBW \cite{GBW} saturation scale 
$k_{sat}=k_0(x_0/x)^{\lambda/2}$ with parameters $x_0$, $k_0$, $\lambda$ to be determined by fit. We are aware of the fact that this approach has obvious 
limitations 
since the saturation line is not impact parameter dependent and is not affected by evolution. However, it provides an energy dependent cutoff which 
is easy to be implemented in a Monte Carlo program, and therefore we consider it  as a reasonable starting point for future
investigations.
We applied our prescription to calculate the $F_2$ structure function and we obtained good descriptions of HERA data, both in scenario with and
without saturation, see Fig. \ref{fig:F2sat}. However, the gluon densities which are used in calculation of  the $F_2$ structure function have very different shape 
and they may have impact on exclusive observables even in HERA range.
\vspace{-.9cm}
\section{Impact of saturation on exclusive observables}
\begin{figure}[t!]
\centerline{\epsfig{file=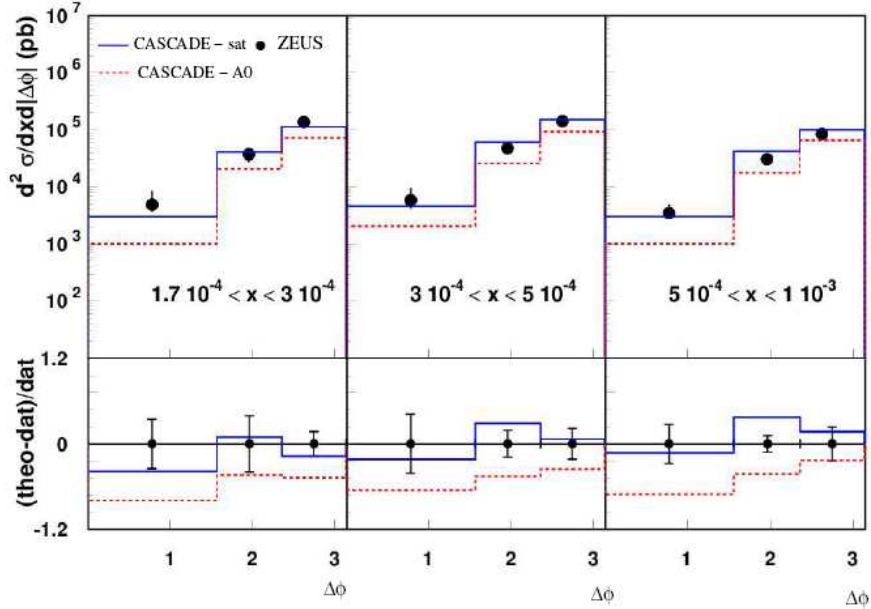,width=12cm}}
\caption{\em (up) Differential cross section for three jet event calculated within CCFM with saturation boundary (blue line) 
compared to CCFM without saturation (red line). (down) Ratio between theory prediction minus data divided by data}
\vspace{-0.1cm}
\label{fig:zeus}
\end{figure}
\begin{figure}[t!]
\centerline{\epsfig{file=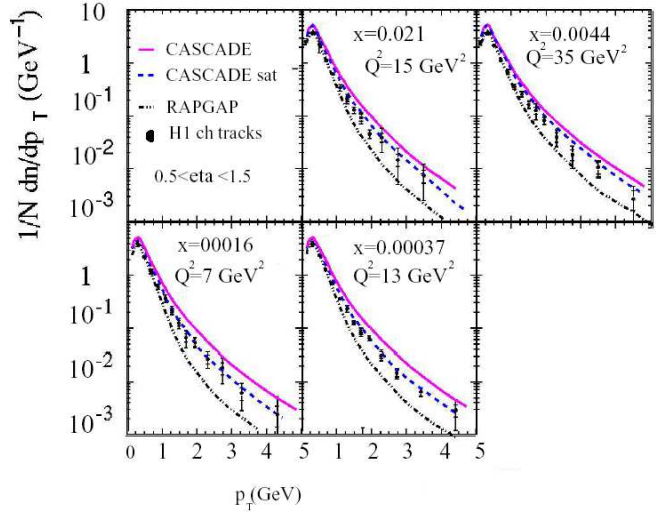,height=7cm}}
\caption{\em Differential cross section for transverse momentum distribution of charged hadrons calculated within CCFM (violet continuous line), 
CCFM with saturation (dashed blue line) and DGLAP (dotted black line)}
\vspace{-0.5cm}
\label{fig:charged}
\end{figure}

Using the gluon density determined by fit to $F_2$ data we may now go on to investigate the impact of saturation on exclusive observables.
As a first exclusive observable we choose the differential cross section for three jet events in DIS \cite{HJ}. Here we are interested in the dependence of
the cross section on the azimuthal angle $\Delta\phi$ between the two hardest jets.
This calculation is motivated by the fact that the produced hard jets are directly 
sensitive to momentum of the incoming gluon and therefore are sensitive to the
gluon  $k_T$ spectrum. In the results we see a clear difference between 
the approach which includes saturation and the one which does not include
it. The description with saturation is closer to data suggesting the need for 
saturation effects.  
Another observable we choose is the $p_T$ spectrum of produced charged particles in DIS.
\cite{H1coll}. 
We compare our calculation with calculation
based on CCFM and on DGLAP evolution equations. From the plots Fig. \ref{fig:charged} we see that the CCFM with saturation describes data better then the
other approaches. CCFM overestimates the cross-section for very low $x$ data
while DGLAP underestimates it. 
This is easy to explain, in CCFM one can get large contributions from larger momenta in the chain due to lack of ordering in $k_T$ while 
 in DGLAP large $k_T$ in the chain is suppressed. On the other hand CCFM with
 saturation becomes ordered for small $x$ both in $k_T$ and rapidity and therefore
interpolates between these two. 

\vspace{-0.5cm}
\section{Conclusions}
In this contribution we studied saturation effects in exclusive observables
using a Monte Carlo event generator. 
Including saturation effects we obtained a reasonably good description of DIS data for $\Delta\phi$ distribution of jets Fig. 
\ref{fig:zeus} and $p_T$ spectrum of produced charged hadrons Fig. \ref{fig:charged}. 
We compared prediction based on an approach with saturation to one which does
not include it, and we clearly see that the approach
based on saturation gives a better description of the measurements. 
\vspace{-0.5cm}
\section*{Acknowledgments}
We would like to thank  E.Avsar, G.Gustafson, Al Mueller for useful discussions.
Useful comments by  G. Gustafson on the manuscript are kindly acknowledged.
\vspace{-0.5cm}
\begin{footnotesize}

\end{footnotesize}
\end{document}